\documentclass[12pt, draftclsnofoot, onecolumn]{IEEEtran}
\usepackage{etex}
\usepackage[font=footnotesize,caption=false]{subfig} 
\usepackage[dvipsnames,svgnames,usenames]{xcolor}
\usepackage{array}
\usepackage{graphicx}
\usepackage[T1]{fontenc}
\usepackage{textcomp}
\usepackage[utf8]{inputenc}
\usepackage[final]{microtype}
\usepackage{icomma}
\usepackage{xspace}
\usepackage[tbtags]{amsmath}
\usepackage{amssymb,amsfonts,bm}
\usepackage{mathtools} 
\usepackage{calc}
\usepackage{ifthen}
\usepackage[noadjust]{cite}
\usepackage{tikz}
\usepackage{etoolbox}
\usepackage{everypage}

\allowdisplaybreaks[3]

\ifCLASSOPTIONdraftcls
\AtBeginEnvironment{figure}{}
\fi

\AddEverypageHook{
\tikz[remember picture,overlay,every node/.style={font={\fontsize{7}{8}\selectfont},align=center}]{
\node [anchor=south] at (current page.south) {\textcopyright 2023 IEEE. Personal use is permitted, but republication/redistribution requires IEEE permission.See https://www.ieee.org/publications/rights/index.html for more information.};
\node [anchor=north] at (current page.north) {This article has been accepted for publication in IEEE Network. This is the author's version which has not been fully edited and content may change prior to final publication.};
}}

\begin{document}
\bstctlcite{IEEEexample:BSTcontrol}
\title{Federated Learning in Satellite Constellations}

\author{
	Bho~Matthiesen,~\IEEEmembership{Member,~IEEE},
	Nasrin~Razmi,~\IEEEmembership{Student Member,~IEEE},\\
	Israel Leyva-Mayorga,~\IEEEmembership{Member,~IEEE},
	Armin~Dekorsy,~\IEEEmembership{Senior~Member,~IEEE},
	and~Petar~Popovski,~\IEEEmembership{Fellow,~IEEE}%
		\thanks{
			B.~Matthiesen, N.~Razmi, and A.~Dekorsy are with the Gauss-Olbers Center, c/o University of Bremen, and the Department of Communications Engineering, University of Bremen, 28359 Bremen, Germany (e-mail: \{matthiesen, razmi, dekorsy\}@ant.uni-bremen.de).
			I.~Leyva-Mayorga and P.~Popovski are with the Department of Electronic Systems, Aalborg University, 9100 Aalborg, Denmark (e-mail: \{ilm, petarp\}@es.aau.dk). P.~Popovski is also holder of the U~Bremen Excellence Chair in the Department of Communications Engineering, University of Bremen, 28359 Bremen, Germany.
		}%
		\thanks{
			This work is supported in part by the German Research Foundation (DFG)
			under Germany's Excellence Strategy (EXC 2077 at University of Bremen, University Allowance).
		}
	}

\maketitle
\thispagestyle{empty}

\begin{abstract}
	Federated learning (FL) has recently emerged as a distributed machine learning paradigm for systems with limited and intermittent connectivity. This paper presents the new context brought to FL by satellite constellations, where the connectivity patterns are significantly different from the ones observed in conventional terrestrial FL. The focus is on large constellations in low Earth orbit (LEO), where each satellites participates in a data-driven FL task using a locally stored dataset. This scenario is motivated by the trend towards mega constellations of interconnected small satellites in LEO and the integration of artificial intelligence in satellites. We propose a classification of satellite FL based on the communication capabilities of the satellites, the constellation design, and the location of the parameter server. A comprehensive overview of the current state-of-the-art in this field is provided and the unique challenges and opportunities of satellite FL are discussed. Finally, we outline several open research directions for FL in satellite constellations and present some future perspectives on this topic.
\end{abstract}

\section*{Introduction}
A rapidly increasing number of satellites is orbiting Earth and collecting massive amounts of data.
This data can be utilized in three principal ways: to train a machine learning (ML) model, to do inference, or to store it for later retrieval.
We are interested in the first use case, where
the conventional approach is to store the data locally and later aggregate it in a centralized location. But this requires massive amounts of bandwidth and storage, induces large delay and energy costs, and potentially infringes upon data ownership.
Instead, recent developments strive to alleviate these downsides and bring the training process towards the data stored distributedly within the satellite constellation.
This requires performing distributed ML (DML) within the constellation, where each satellite uses its collected data to locally train an ML model.
The DML paradigm that operates in the context of systems with limited and intermittent connectivity is known as federated learning (FL) \cite{McMahan2017,FLopenNow}. There, the clients work independently on ML model updates based on their local data and share intermediate results among each other using a central parameter server (PS). Communication with this PS is an integral part of the training process and developing efficient communication protocols to support FL is recognized as one of the major design challenges for the sixth generation of communication systems \cite{Uusitalo2021}.
Besides the apparent benefits of directly training an ML model close to the data, implementing FL within satellite constellations is also an important step towards seamless integration of terrestrial and non-terrestrial networks \cite{Chen2022}.

\begin{figure}[t]
\centering
\subfloat[]{\includegraphics{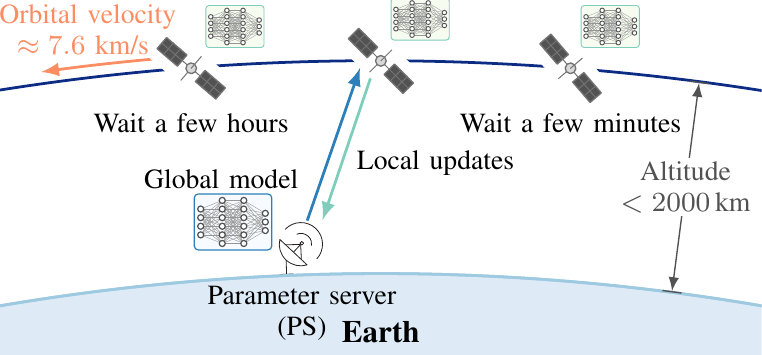}}\hfil
\subfloat[]{\includegraphics{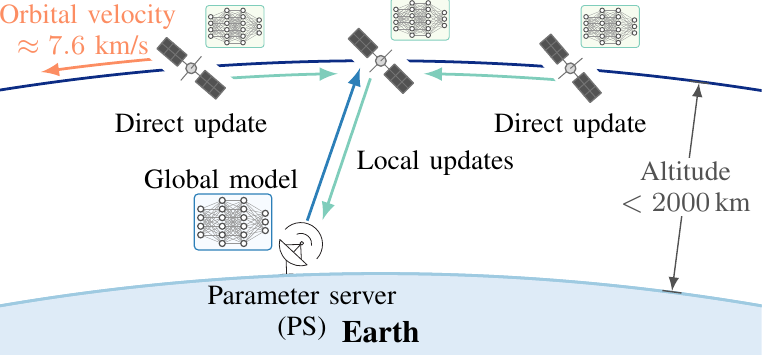}}
\caption{FL in a satellite constellation (a) without and (b) with ISLs. In the former case, the satellites must wait until the next pass to send their local update and to receive the updated global model. In the latter case, the satellites can update the models if at least one of them has connection to the PS. In both cases, the connectivity of the satellites to the PS is predictable.}
\label{fig:sysmod}
\end{figure}

The goal of this article is to analyze the challenges and opportunities of in-constellation training of ML models from data, collected by a satellite constellation in a distributed manner.
We focus on large constellations in low Earth orbit (LEO), where each satellite participates in a data-driven FL task using its local data set. 
A PS orchestrates this process from outside the constellation, located either in a ground station (GS) or in a satellite with larger orbital height.
This scenario is illustrated in Fig.~\ref{fig:sysmod}. It is motivated by the recent trend towards mega constellations of interconnected small satellites in LEO \cite{satmagazine} and efforts to deploy artificial intelligence (AI) in satellites \cite{Giuffrida2021}. It essentially provides the infrastructure for (re-)training DML models from data stored within the satellites.
Potential applications include improving AI-based communication and network modules \cite{Vazquez2021a,Tang2022}, Earth observation missions \cite{Izzo2022}, and constellation management.

Compared to conventional FL in mobile networks, the spatio-temporal scope of FL is significantly augmented in orbital networks. This is because the movement of satellites is completely determined by the laws of orbital mechanics.
A closely related scenario is FL in unmanned aerial vehicle (UAV) networks \cite{Qu2021}. However, there are subtle but significant differences. While the trajectories of UAVs can be controlled during deployment, the orbital parameters of satellites are mostly fixed. Moreover, satellites usually cover much larger areas than UAVs, might be unavailable regularly for extended periods of time, and are subject to higher transmission delays due to the long distances involved. From a networking perspective, the \emph{a priori} knowledge of the exact trajectories allows for highly specialized communication schemes to best support the DML process. Instead, UAV networks require more flexible communication approaches that rely on active control and optimization during the learning process.

System architectures for satellite-assisted training of ML models are presented in \cite{Chen2022}. The authors focus on the interaction of remotely located terrestrial devices with a cloud server, assisted by LEO satellites. While this includes a scenario where ML training is conducted on a satellite, the scope is on comparing different cloud-connectivity architectures. Instead, we consider the interaction among satellites within a constellation to distributedly train a ML model. Indeed, while
satellite FL was first proposed in \cite{Razmi2022}, this paper is the first top-level study on orchestrating FL in satellite constellations.
The communication and computation approach to satellite FL depends crucially on the satellites' communication capabilities, the constellation design, and the location of the PS.
We propose a classification of satellite FL based on these aspects and the resulting connectivity patterns.
We show that three classes are sufficient to cover all relevant satellite FL scenarios. In addition, we also show that these three classes are necessary and significant, as the technical challenges are very different for each. This is one of the key contributions of this paper.
We review the state-of-the-art within this framework, discuss promising solution approaches for each scenario, and highlight open topics and future directions for research.

\section*{Fundamentals}
\subsection*{Distributed Machine Learning}
The traditional domain of DML training are data centers with homogeneous client nodes, constant availability, and high bandwidth. This facilitates frequent synchronization between nodes and the potential to (re-)distribute the training data as needed. However, these conditions are not met within the context of mobile (and satellite) networks, where the clients are subject to intermittent availability and limited connectivity. This modified setting motivated the development of federated learning \cite{McMahan2017,FLopenNow}, where
multiple devices collaborate to solve an ML problem using their locally stored data under the coordination of a central PS.
Crucial aspects of FL are that each client's raw data is kept local and not shared with other clients, and that concessions towards bandwidth efficiency are made. Further differentiating factors of FL over DML are non-IID and unbalanced data distribution, systems heterogeneity, and the data being distributed over a massive number of clients \cite{McMahan2017,FLopenNow}.

These aspects need to be addressed by the training procedure, which is essentially a distributed optimization process that minimizes a cost function  based on the local data set with respect to the ML model parameters.
 This is an iterative process operating over a number of rounds, where, in each round, the PS samples a subset of clients and broadcasts the current global model to them. Each client then performs several local gradient steps using a gradient-based optimizer such as stochastic gradient descent. Based on this, the difference between the updated local model and the current version of the global model is computed and
communicated to the PS. The PS waits for the reception of the local updates from all scheduled clients, aggregates and applies them in a gradient step to obtain a new iterate of the global model, and starts a new round until convergence.
This procedure is known as federated averaging (FedAvg) \cite{McMahan2017}.

\subsection*{Synchronization in Parallel Algorithms}
Coordination is an integral part of every distributed procedure. A common approach is to divide a parallel algorithm into phases, during which each client performs local computations based on the results from previous phases. All communication between clients is performed at the end of each phase. Since this includes sharing the results from the current phase, a new phase can only start once all clients complete their computations.
Algorithms operating according to this scheme are called \emph{synchronous}. For example, the FedAvg algorithm and its descendants belong to this class.

Waiting for all clients to complete the current phase can incur substantial delays. This synchronization penalty might be averted by using an asynchronous algorithm with less strict coordination. There, the communication phases of the clients are not synchronized and, hence, local computations might rely on stale results from other clients. While asynchronous algorithms are not subject to synchronization delays, the stale data might lead to slower convergence speed. This trade-off always needs to be considered when deciding on the appropriate coordination approach for a distributed algorithm. Instances of asynchronous FL methods are FedAsync \cite{Xie2020} and FedSat \cite{Razmi2022}.

\subsection*{Satellite Constellations}
Satellites are classified based on their altitude of deployment, which determines their orbital characteristics. LEO satellites are non-geostationary satellites in altitudes ranging from 500\,km to 2000\,km, below medium Earth orbit (MEO). Contrary to geostationary (GEO) satellites, these require a high orbital velocity to maintain their orbit and, hence, move rapidly with respect to Earth's surface. Since, the orbits of LEO satellites do not compensate for the rotation of the Earth, every satellite pass might occur at a different position with respect to the ground infrastructure. This creates the need for continuous adaptation and management of radio resources.

\begin{figure}
    \centering
    \includegraphics{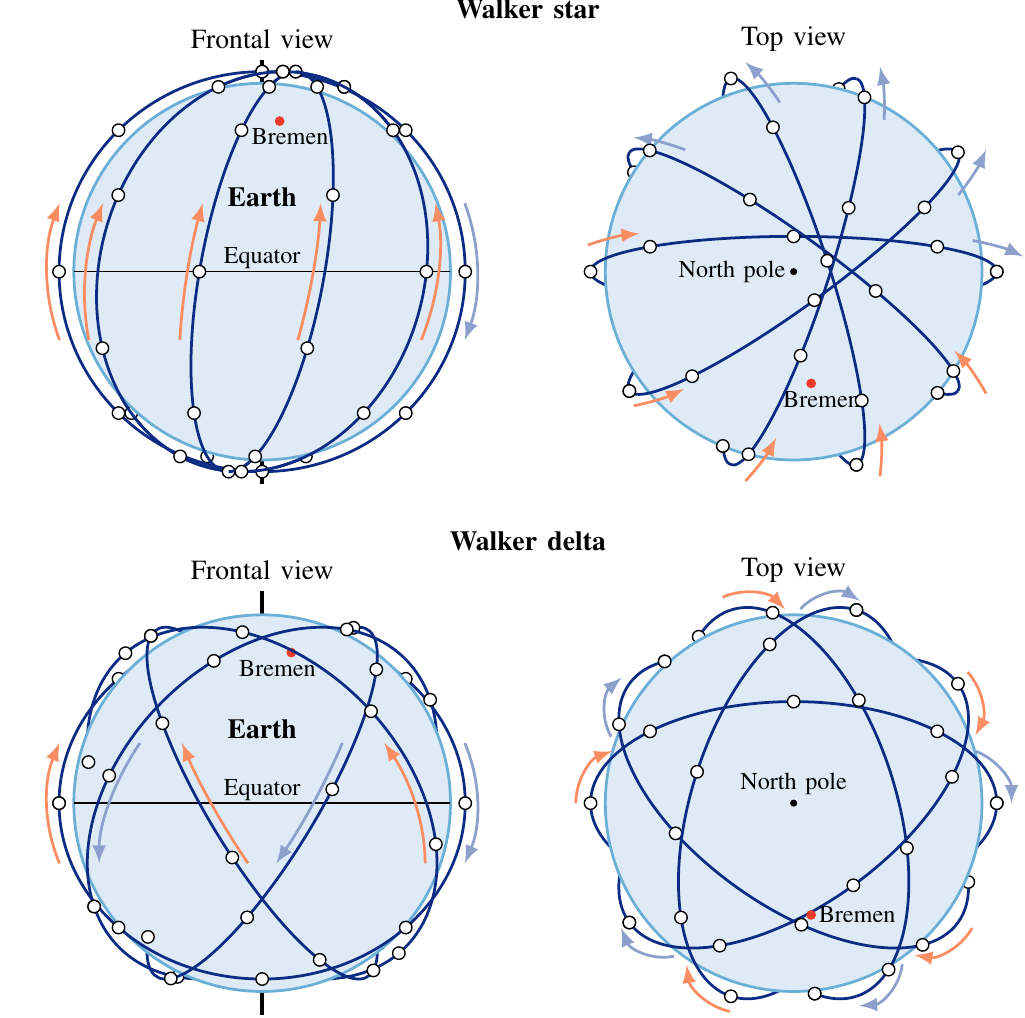}
\caption{Illustration of a 80\textdegree: 40/5/1 Walker star and a 60\textdegree: 40/5/1 Walker delta constellation, both at an altitude of 500\,km. The frontal view is for an observer in the equatorial plane at 0\textdegree{} longitude and the top view from the polar plane towards the North pole with 0\textdegree{} longitude pointing down.}
    \label{fig:constellations}
\end{figure}

Constellations are groups of satellites working together as a single system. Satellites within a constellation are typically organized in orbital planes, which are groups of satellites with similar altitude of deployment and orbital \emph{inclination}. The inclination of the orbital plane determines its alignment with the equatorial plane, where an orbit with inclination 0\textdegree{} would be fully aligned with the equatorial plane, and a polar orbit has an inclination of 90\textdegree.
Two of the most common satellite constellations are Walker star and Walker delta, illustrated in Fig.~\ref{fig:constellations}.
Both are  defined by the notation $i\!:\!t/p/f$, where $i$ is the inclination, $t$ the total number of satellites, and $p$ the number of evenly spaced orbital planes. The phase parameter $f$ is used to determine the relative spacing between satellites in adjacent orbital planes, which are shifted along their orbit by $360^\circ f/t$ \cite{SatBook}.

\subsection*{Satellite Communications}
 Maintaining a communication link towards the ground infrastructure or to a satellite in a higher orbit is a baseline functionality. Typical contact times between LEO satellites and GSs are in the order of a few minutes. In modern satellites, these links are often complemented by inter-satellite communication links, which come in two flavors: intra- and inter-orbit links \cite{satmagazine}.
 
Intra-orbit links connect adjacent satellites within the same orbital plane, effectively forming a ring network. Due to the stable relative position between satellites within the same orbit, intra-orbit communication links have nearly constant channel characteristics and can be modelled as fixed rate bit-pipes.

Instead, inter-orbit links connect satellites across orbital planes and are more heterogeneous in nature. As shown in Fig.~\ref{fig:constellations}, in Walker star constellations, satellites from adjacent orbital planes travel in similar directions and can maintain communication links with time-varying distance. However, the same constellation also contains satellites traveling in nearly-opposite directions, that is, with high relative velocity that leads to transmission impairments and short contact times~\cite{satmagazine}.

\section*{A Classification of Satellite FL Scenarios}

Connectivity between clients and PS is primarily determined by orbital mechanics and the satellites' communication capabilities. Combined with the reality of satellite constellations being owned and operated by a single entity, the spatio-temporal scope of FL is significantly altered from stochastic towards deterministic and predictable client availability.
Another aspect is that, in satellite FL, the number of clients participating in the learning process is orders of magnitude smaller compared to conventional FL, resulting in each client having a relatively larger share of the available data and playing a more influential role in the learning process.
Therefore, it appears prudent to strive for including as many devices as possible in each round instead of stochastically sampling over the devices.

The deterministic, predictable, and partially controllable device participation is the key factor that differentiates satellite FL from the conventional scenario. While these features are beneficial for orchestrating the learning process, they also come at the cost of orbital mechanics dictating the connectivity between clients and the PS.
Indeed, limited communication opportunities are often the primary impairment to the convergence speed in satellite FL.
Following this observation, satellite FL scenarios can be divided into three classes based on the connectivity patterns between satellites and the PS. Each class requires a distinct approach to achieve the fastest possible convergence speed given the satellites' capabilities.

In order to clearly define these classes, it is useful to introduce a graph-theoretic model to represent the underlying network.
Its spatio-temporal relationships are best captured by a temporal graph with static vertex set and a set of time-edges connecting these vertices at specific points in time.
At each time instance, we can take a snapshot of this temporal graph to obtain a conventional static  graph representing the current spatio-temporal state of the network. Each communication node within this network, that is, satellites and GSs, is represented as a vertex. Communication between two nodes is feasible at a specific time instance if there exists a time-edge between the corresponding vertices for the complete duration of the communication, including the propagation delay between those nodes.
A notable exception from the general spatio-temporal properties of this network are intra-orbit links, which have no time-dependence due to the stable relative position of the satellites. Within the temporal graph describing the network, these nodes form a static subgraph that we denote as a cluster. These subgraphs are usually cyclic and, in most cases, circular, that is, each consists of a single cycle since it models an intra-orbital ring network.

\begin{figure}
	\centering
	\includegraphics{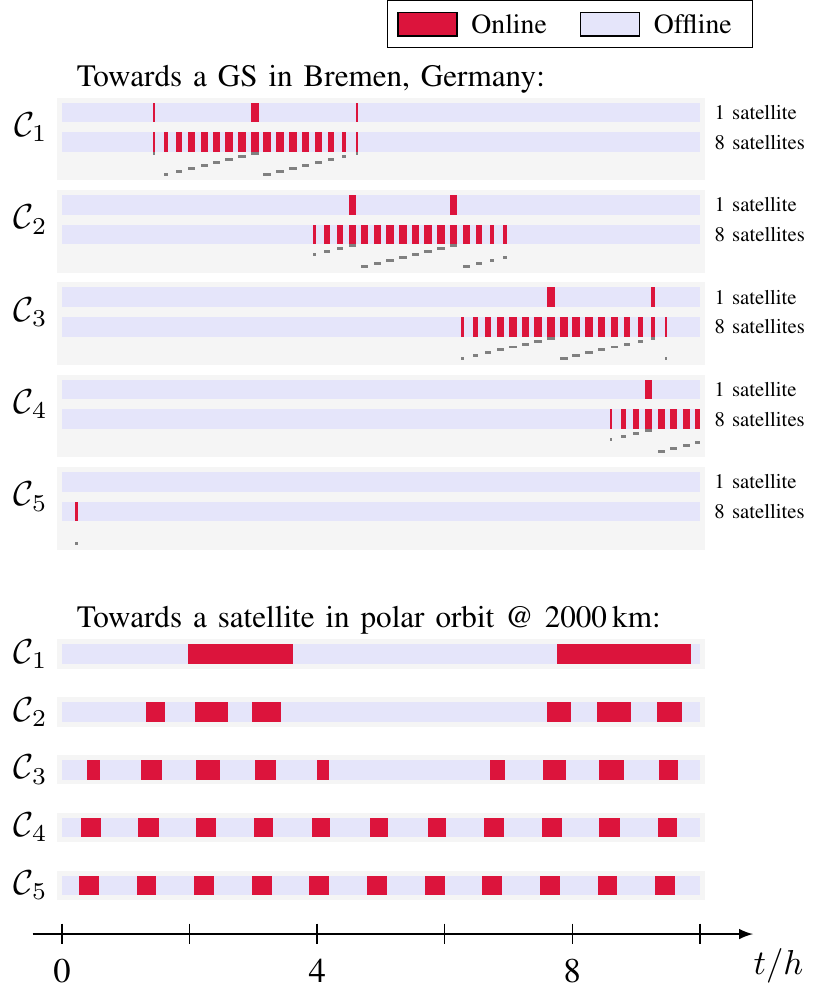}%
\caption{Connectivity pattern from satellites in a Walker star constellation with five orbits at 80\textdegree{} inclination and 500\,km altitude towards a GS in Bremen, Germany, and a satellite in polar orbit at 2000\,km altitude, respectively. In addition, cluster connectivity towards the GS is shown for a 80\textdegree: 40/5/1 Walker star constellation, that is, the same constellation as before but with eight satellites per orbit. Connectivity of individual satellites within each orbital plane  towards the GS is shown in grey below the cluster connectivity. In all cases, the connectivity towards the out-of-constellation node is sporadic.}
	\label{fig:pattern:sporadic}
\end{figure}

\begin{figure}
	\centering
	\includegraphics{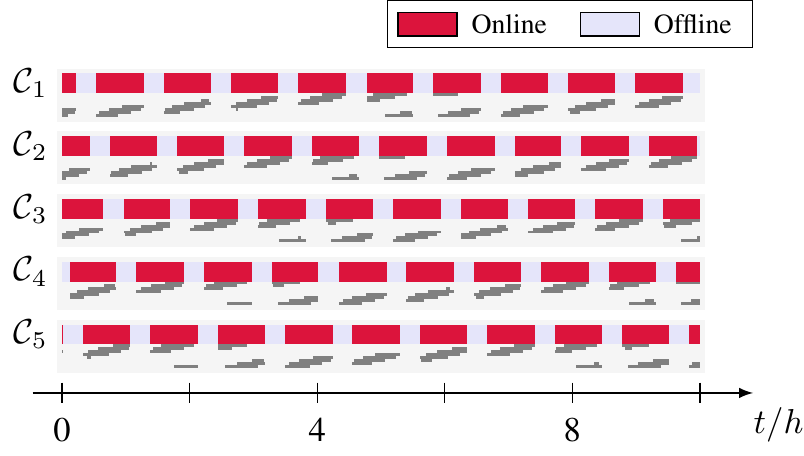}%
\caption{Connectivity pattern from five client clusters towards a satellite in equatorial orbit at 2000\,km altitude. Each client cluster corresponds to an orbital plane in a 80\textdegree: 40/5/1 Walker star constellation where satellites are equipped with intra-orbit ISLs. Connectivity of individual satellites within each orbital plane is shown in grey below the cluster connectivity. While the cluster connectivity towards the out-of-constellation satellite is close to persistent, the connectivity of the individual satellites is sporadic.}
	\label{fig:pattern:dense}
\end{figure}

An important characteristic of clusters is the stable inter-satellite connectivity, which enables the expansion of the communication window between the PS and individual clients through multi-hop routing. This means that every client within a cluster can communicate with the PS as long as there is at least one client in the cluster with a direct connection to the PS. Hence, we base the classification of satellite FL on the connectivity patterns among clusters.
Some example connectivity patterns between a PS and five client clusters are displayed in Figs.~\ref{fig:pattern:sporadic} and~\ref{fig:pattern:dense}. There, different PS locations are considered and all client satellites are in LEO at 500\,km altitude, organized in a Walker star constellation with five orbital planes at 80\textdegree{} inclination. Either a single or eight satellites per orbital plane are simulated and each satellite is connected to its direct orbital neighbors through an inter-satellite link (ISL). Thus, these satellites form single cluster. Time intervals, where a connection between at least one node within the cluster and the PS is possible, are marked in red, while individual direct per-satellite connectivity towards the PS is displayed in dark gray below the cluster connectivity.

Based on these connectivity patterns, satellite FL can be categorized into three classes.
\begin{itemize}
	\item \textbf{Sporadic direct connection to PS,} where communication between a client cluster and the PS is only possible for short time intervals with long offline periods in between. This is the most common scenario in satellite FL when there is no inter-satellite connectivity.  Example connectivity patterns are displayed in Fig.~\ref{fig:pattern:sporadic} for cluster connectivity towards a PS located in a GS in Bremen, Germany, and a LEO satellite in a 2000\,km polar orbit. The core challenge in this scenario is to adapt and organize the learning process such that every communication opportunity is utilized to maximize convergence speed and accuracy.

	\item \textbf{Near-persistent direct connection to PS,} where a connection between the PS and at least one client per cluster is possible most of the time. This is a model for many constellations with intra-orbit ISLs. For example, Fig.~\ref{fig:pattern:dense} displays the connectivity pattern of each orbital plane in a 80\textdegree: 40/5/1 Walker star constellation at 500\,km towards a satellite in equatorial orbit at 2000\,km.
		The primary technical challenges in this class are related to networking, particularly in the areas of routing and in-network computation. For maximum efficiency, these techniques should be tailored for FL and leverage the predictable movement of satellites.
	\item \textbf{Multi-hop connection to PS via inter-cluster connectivity,} where clusters can communicate among each other, that is, satellites have inter-orbit communication capabilities.
	Among the considered scenarios, this is the most advanced and challenging, but also that which offers the greatest potential for efficient communication techniques and exploiting the unique properties of satellite FL.
\end{itemize}

\begin{figure}
	\centering%
	\includegraphics{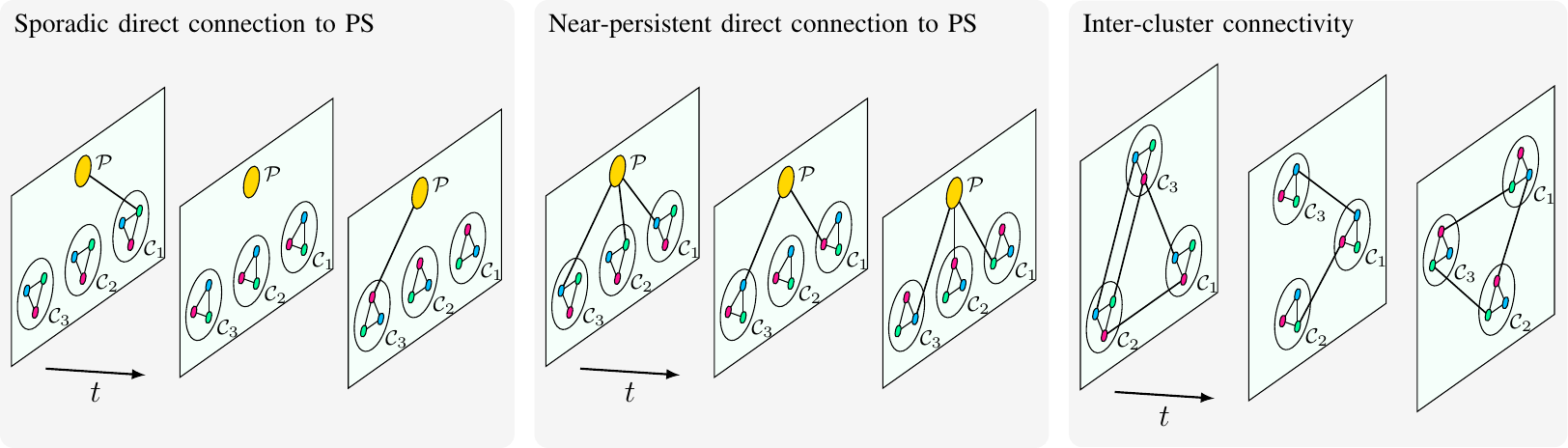}%
	\caption{Spatio-temporal graph models for FL in satellite constellations for different connectivity patterns. Three snapshots of the time graph are displayed for each scenario. Client clusters are denoted as $\mathcal C_i$, while the PS is labeled as $\mathcal P$.}
	\label{fig:graphs}
\end{figure}

Spatio-temporal graph models for these three scenarios are illustrated in Fig.~\ref{fig:graphs}, where three snapshots of the time graph are shown for each scenario. The client clusters and PS are denoted as $\mathcal C_i$ and $\mathcal P$, respectively. Observe that the third graph does not contain a PS, since the distinction in client and PS cluster is not necessary in this case.
The communication and computation approaches for each class differ substantially. In the remaining part of this article, we discuss the challenges and opportunities of each scenario.

\section*{Sporadic Direct Connection to PS}
In the sporadic connectivity scenario, the  clients have long offline periods between comparably short intervals of connectivity towards the PS. This is usually the case when satellites have no ISLs or the learning process is orchestrated by a single GS (cf.~Fig.~\ref{fig:pattern:sporadic}). Directly applying a synchronous FL algorithm, for example, FedAvg,  would result in a very slow convergence.
To illustrate, consider running FedAvg under full client participation in a network with connectivity pattern as in Fig.~\ref{fig:pattern:sporadic}, with a single satellite per orbit and connectivity towards a GS. Recall that the FedAvg algorithm operates synchronously, that is, it waits for all selected clients to return their model update before aggregating the updates into a new global model. Assuming learning starts at $t = 0$ in Fig.~\ref{fig:pattern:sporadic}, each client obtains the initial version of the global model from the PS upon its first connection opportunity. It then starts computing a local update and transmits it at the second contact to the PS. The PS waits until all updates are received before updating the model. Hence, the clients have to wait at least until their third connection to the PS before receiving the first iteration of the global model. Referring to Fig.~\ref{fig:pattern:sporadic}, it can be observed that, even without $\mathcal C_5$, this first global iteration of FedAvg will take over nine hours to complete due to $\mathcal C_4$.

This problem can be somewhat alleviated by client scheduling at the PS. For example, consider the second connectivity pattern in Fig.~\ref{fig:pattern:sporadic} and assume a new training iteration starts at $t \approx 3\,\mathrm{h}$, when $\mathcal C_2$ connects to the PS to deliver its update. Further assume that the satellites require approximately 30\,min for local computation. The scheduler
must make a choice among several ``sub-optimal'' options. Either it schedules
$\mathcal C_3$ to $\mathcal C_5$, which will lead to the next global iteration being finished around $t \approx 4\,\mathrm{h}$. This will leave $\mathcal C_1$ and $\mathcal C_2$ idle. The scheduler can also include $\mathcal C_1$ and hope that it will deliver its update before going offline for several hours. If it does not manage to complete within that time, the PS can either neglect that update and continue, or wait for $\mathcal C_1$ to come online again. The first option leads to wasting energy and computational resources in $\mathcal C_1$, the second leads to a significant delay in the global FL process with $\mathcal C_3$ to $\mathcal C_5$ being idle. If $\mathcal C_1$ manages to complete in time, it will still be idle during its long offline period because the current iteration will not be finished before it disconnects from the PS. No matter what the scheduler decides, either $\mathcal C_2$ will be idle or $\mathcal C_4$ and $\mathcal C_5$, leading to unused computational resources and, thus, considerable delay in the training process.

\subsection*{Asynchronous Federated Learning}
A completely different strategy is using an asynchronous FL algorithm,
which avoids missing communication opportunities altogether.
Recall that synchronous FL methods like FedAvg wait for all selected clients to transmit their results before updating the global model and starting a new iteration. Instead, an asynchronous algorithm updates the global model whenever a client update arrives at the PS.
This allows the clients to constantly work on local updates while being offline. An asynchronous version of FedAvg designed with satellite FL in mind is developed in \cite{Razmi2022} and named \emph{FedSat}. The core idea is to implement the PS update rule of FedAvg \emph{incrementally} and have satellites always work on computing a new update to the model.

\begin{figure}
	\centering
	\includegraphics{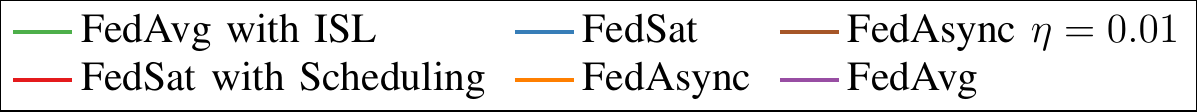}\\[0.6em]
	\subfloat[PS located in a GS in Bremen, Germany. No inter-satellite connectivity.]{%
		\includegraphics{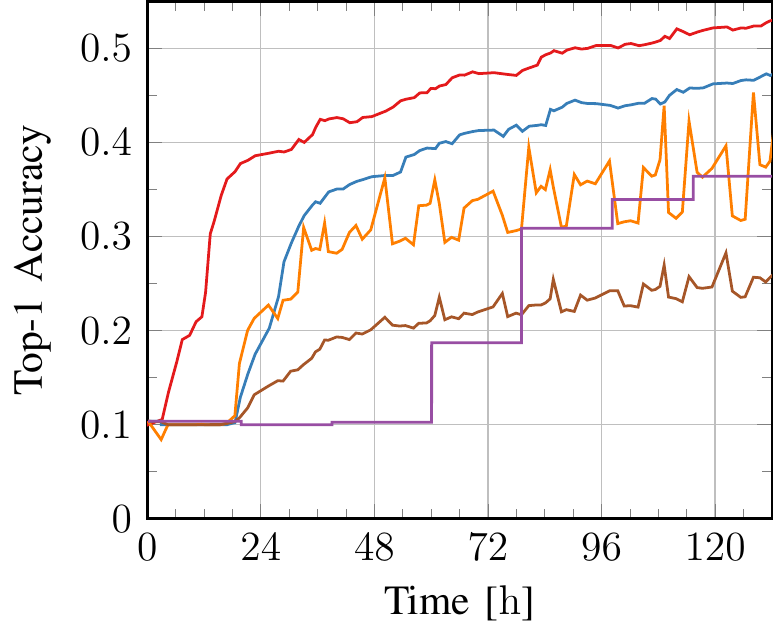}%
		\label{fig:async:conv}%
	}
	\hfill%
	\subfloat[PS located in a satellite in equatorial orbit at 2000\,km altitude.]{%
		\includegraphics{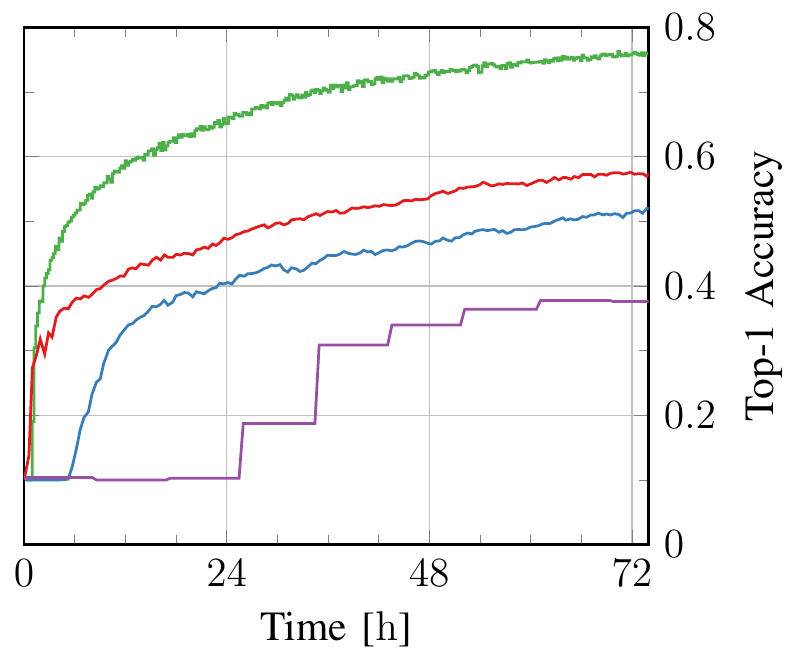}%
		\label{fig:isl:conv}%
	}%
\caption{FL in a 80\textdegree: 40/5/1 Walker star constellation. Top-1 accuracy as a function of the (simulated) wall clock time for training a ResNet-18 with non-IID CIFAR-10 data. Training is conducted in batches of size 100 with a learning rate $\eta = 0.1$. Computation of a local model update at the satellites is assumed to require 1\,min.}
\end{figure}

A comparison of FedSat with FedAvg and FedAsync \cite{Xie2020}, a general purpose asynchronous FL method, in terms of convergence time is shown in Fig.~\ref{fig:async:conv}. There, a ResNet-18 is trained on the CIFAR-10 data set in a 80\textdegree: 40/5/1 Walker star constellation at 500\,km altitude.
This is a classification task where images are categorized into ten different classes, for example, as an airplane, ship, or truck.
A non-IID distribution of the training data is considered, where half of the classes is distributed equally among the satellites in orbits one, two, and the first four satellites in orbit three. The second half of the classes is distributed equally among the remaining 20 satellites.
Learning is orchestrated by a PS located in a GS in Bremen, Germany. It can be observed from Fig.~\ref{fig:async:conv} that FedSat \cite{Razmi2022} clearly outperforms the state-of-the-art approaches.

Convergence of FedSat is established in \cite{Razmi2022} under the assumption that all satellites have the same altitude. This assumption is required to obtain equal asymptotic participation rates for all satellites. Applying FedSat in scenarios with different orbital heights  might result in a biased solution and, hence, reduced test accuracy. However, this issue is common to most FL algorithms, including FedAvg and FedAsync. The numerical results in \cite{Razmi2022} clearly establish that FedSat outperforms these algorithms even in scenarios with multiple orbital shells and non-IID data distribution. Still, properly addressing varying participation rates algorithmically remains an open topic for future research.
Further open topics are combining asynchronous FL with the intra-cluster communication techniques discussed for the near-persistent connectivity scenario, and hierarchical asynchronous FL to facilitate aggregation through multiple GS.

\subsection*{Predictive Scheduling for FedSat}
A major impairment to convergence speed in asynchronous FL is model staleness, that is, local updates delivered to the PS can be based on significantly outdated versions of the current global model. Gradient steps computed on such an outdated version of the global model parameters are potentially misaligned with the current gradient and, thus, do not necessarily improve the objective function. A straightforward scheduling solution to address this issue relies on the predictable movement of satellites \cite{nasrinEusipco2022}. The idea, for each satellite during its contact to the PS, is to estimate whether a local training iteration can be completed \emph{during the next contact}. If this is the case, the satellite does not retrieve the global model and remains idle while being offline. Upon the next contact, it will obtain the current global model and complete a local iteration while being visible to the PS. Thus, the update will be based on a much more recent version of the global model, which leads to improved convergence speed. This can be observed from Fig.~\ref{fig:async:conv}, where FedSat in combination with scheduling converges significantly faster than the algorithms without.
Future opportunities for research include more elaborate scheduling approaches that, for example, take the energy consumption into account.

\section*{Near-Persistent Direct Connection to PS}
In this scenario, at least one node per client cluster can connect to the PS most of the time. This permits the efficient usage of a synchronous FL procedure like FedAvg. 

With only one satellite per cluster, this is the conventional FL setup. However, except for LEO constellations with a PS in a high MEO or GEO, setups with singleton client clusters usually fall within the sporadic PS connectivity scenario. Thus, we focus on the case with multi-satellite clusters, where each cluster's satellites are 
in the same orbital plane and connected via ISLs.

The main idea is, for each cluster, to route all client updates in a synchronized fashion over the node currently connected to the PS.
This not only enables synchronous FL without extensive delays but also has the potential to reduce the communication load on the PS.
Due to propagation and processing delays, satellites connected to the PS at the start of the transmission are unlikely to be connected to it after a message has been forwarded over multiple hops. Hence, route computation needs to be predictive and should actively exploit the determinism of the underlying temporal network.
The straightforward approach is to forward each local update individually and use contact graph routing. However, the synchronous FL PS is not interested in individual client updates but in a combination of them, for example, the weighted sum for FedAvg. By computing this combination incrementally using in-network computing, the communication load in the network can be reduced considerably. In particular, each node within a cluster waits for all scheduled incoming local updates it has to forward, and computes a partial aggregate from the received and its locally computed update. This \emph{partial aggregate} is then forwarded to the next hop towards the PS and keeps the communication effort of each node at the size of a single model update. Instead, with conventional unicast forwarding, the number of model update transmissions scales quadratically with the cluster size.

Implementing this \emph{incremental aggregation} protocol requires a certain amount of coordination among the satellites, because every satellite needs to know whether it has to wait for incoming updates and where it has to forward them to. This is achieved with remarkable simplicity since, unlike common FL, the clients are all under central control and the orbital positions of all involved nodes are completely predictable. Upon distribution of a new global model, which initiates a new training round, each client cluster determines a satellite that \emph{will be} in a favorable position to communicate with the PS when all local updates from the cluster's nodes  are expected to arrive at that satellite. Then, a shortest path tree towards the selected satellite is computed and communicated to all nodes. This \emph{predictive routing} process is based on models for computation, communication and processing delays. 
The complete protocol is described and analyzed in \cite{nasrinICC2022}.

In Fig.~\ref{fig:isl:conv}, the improvement in convergence time due to using intra-orbit ISLs is evaluated with respect to the algorithms described in the previous section. As before, a 80\textdegree:~40/5/1 Walker star constellation at 500\,km altitude is considered and a ResNet-18 is trained on the CIFAR-10 data set with non-IID distribution. This time, the PS is located in a LEO satellite in equatorial orbit at 2000\,km altitude, which corresponds to the connectivity pattern in Fig.~\ref{fig:pattern:dense}.
The gain of using ISLs is tremendous, which is mainly due to the increased number of global training iterations manageable within the same time span and due to stronger convergence properties of synchronous algorithms. Moreover, the communication effort at the PS is reduced from receiving 40 individual model updates to only five partial aggregates per global iteration.

The algorithm proposed in \cite{nasrinICC2022} shows promising results, but operates under simplified deterministic models for communication and computation. More robust routing and error handling algorithms are required to adapt this approach to realistic satellite operation conditions. Especially the combination of incremental aggregation with common gradient compression techniques for FL will have considerable impact on the underlying timing and, thus, on the predictive routing procedure.

\section*{Multi-hop Connection to PS via Inter-Cluster Connectivity}
This scenario is concerned with connectivity among arbitrary clusters, instead of limiting the communication to links between individual clusters and the PS. The main complication is that routes, in general, now incorporate multiple hops over time-edges instead of just having a single final time-edge hop. Sensible assumptions motivated by the realities of satellite constellations are that connections between clusters are time-edges that can have relatively short lifespan and that the network graph is connected for all time instances.
The communication and computation approach is again to use predictive routing, incremental aggregation, and synchronous FL as in the previous scenario.

Multiple hops over time-edges in combination with predictive routing leads to a considerably higher routing failure probability than before, since errors are likely anywhere within the route and not only at the last hop. This calls for sophisticated failure handling protocols and poses the question whether a completely decentralized routing approach would be better suited. Another option to add resilience and reduce routing failures is multipath routing, where model updates from individual clients arrive over multiple paths at the PS. However, all these approaches complicate incremental aggregation considerably as they might lead to multiple copies of the same local model update being incorporated in the partial aggregates arriving at the PS. Overcoming these inhibitions appears difficult but will enable true in-constellation FL without the need for external orchestration.

\section*{Conclusions}
We have introduced and characterized satellite FL, which  differs substantially from conventional terrestrial FL.
A fundamental observation is that there does not exist a \emph{single} satellite FL scenario but three distinct cases. We have identified and defined these classes based on the
satellite's communication capabilities and the resulting connectivity patterns towards the PS. For each of these classes, we have reviewed the state-of-the-art, discussed the principal networking and federated optimization challenges, and highlighted future directions for research.
Especially the third class, which requires inter-orbit links, is largely unexplored and offers significant potential for innovation.

\bibliography{IEEEabrv,IEEEtrancfg,references.bib}

\vspace*{-2\baselineskip}
\begin{IEEEbiographynophoto}{Bho Matthiesen} [M] (matthiesen@uni-bremen.de)
	is a research group leader at the University of Bremen.
	He is an Associate Editor of EURASIP Journal on Wireless Communications and Networking and an editorial board member for Scientific Reports (Nature Portfolio). His research interests are in communication theory and  wireless communications.
\end{IEEEbiographynophoto}
\vspace*{-3\baselineskip}
\begin{IEEEbiographynophoto}{Nasrin Razmi} [S] (razmi@ant.uni-bremen.de)
	received the B.Sc.\ degree in electrical engineering from Urmia University and the M.Sc.\ degree from the K.\,N.~Toosi University of Technology (KNTU). She is currently pursuing the Ph.D.\ degree at the University of Bremen, Bremen, Germany. Her research interests include wireless communication and federated learning.
\end{IEEEbiographynophoto}
\vspace*{-3\baselineskip}
\begin{IEEEbiographynophoto}{Israel Leyva-Mayorga} [M] (ilm@es.aau.dk)
	received the Ph.D.\ degree (cum laude and extraordinary prize) in telecommunications from the Universitat Polit\`{e}cnica de Val\`{e}ncia (UPV), Spain, in 2018. He is currently an Assistant Professor at Aalborg University (AAU), Denmark. He is an Associate Editor for IEEE Wireless Communications Letters.
\end{IEEEbiographynophoto}
\vspace*{-3\baselineskip}
\begin{IEEEbiographynophoto}{Armin Dekorsy} [SM'18] (dekorsy@ant.uni-bremen.de)
	is a Professor at University of Bremen, where he is Director of the Gauss-Olbers Space-Technology Transfer-Center and head of the Department of Communications Engineering. He worked in leading research positions in industry for eleven years. His research interests are in the area of signal processing and wireless communications.
\end{IEEEbiographynophoto}
\vspace*{-3\baselineskip}
\begin{IEEEbiographynophoto}{Petar Popovski} [F'16] (petarp@es.aau.dk)
is a Professor at Aalborg University, where he heads the section on Connectivity and a Visiting Excellence Chair at the University of Bremen. He is the Editor-in-Chief of IEEE JOURNAL ON SELECTED AREAS IN COMMUNICATIONS. His research interests are in the area of wireless communication and communication theory.
\end{IEEEbiographynophoto}

\end{document}